# Distributed Applications in Gamification of the Learning Process

Martin Zagar
*Web and Mobile Computing Department*
*RIT Croatia*
Zagreb, Croatia
martin.zagar@rit.edu

Matija Sipek
*Web and Mobile Computing Department*
*RIT Croatia*
Zagreb, Croatia
mxs1943@rit.edu

Nikola Draskovic
*International Business Department*
*RIT Croatia*
Zagreb, Croatia
nikola.draskovic@croatia.rit.edu

Branko Mihaljevic
*Web and Mobile Computing Department*
*RIT Croatia*
Zagreb, Croatia
branko.mihaljevic@croatia.rit.edu

*Abstract*— **Driven by the fact that many of us experienced softer or not-so-soft lockdown, the intention of a couple of instructors at our university was to develop a collaborative tool that could help in online delivery and gamification on two courses that are delivered in the Business and IT curriculums we are offering to our students. That tool could be described as a decentralized web application that simulates Internet marketing principles and helps in gamification of the learning process for our students. We planned our web application for Internet marketing simulation as the gamification of the learning process, which is one of the basics for active learning for Internet Marketing course for International Business students, to gain new class activities by online simulation competing in the field of Internet marketing principles; and for IT students in developing the Web application and also on adopting Blockchain technologies for the distributed reports which need to have a consensus of all teams included in the simulation. The proposed solution includes the design of business logic simulation and using four main digital marketing tools – social networking, content creating and sharing, search engine marketing, and display advertising in use of such application for hands-on online class exercises.**

## I. Introduction

Studying, planning, and developing distributed applications, differs significantly compared to traditional, centralized software applications, namely managing internal persistent memory and operations conducted with it. From an educational point of view, this process is not covered completely in a proper way. IT students lack knowledge on how to build these specific applications due to a lack of business background and related experience. On the other hand, Business students may have the knowledge of business logic an Internet simulation would apply, but they have no technical knowledge needed for the application development.

For the IT students, the switch from centralized application deployment to a decentralized approach introduces new options regarding different principles in software development where access management, integrity, and immutability are imperative. Business students need to understand Internet marketing principles comprehensively, so there is a need for shifting the educational process, for both IT and Business students. Also, according to the common rationale, a motivating learning environment will result in better-performing students and their higher level of satisfaction with both delivered courses and instructors [1]. Therefore, it is essential for curriculum designers to develop programs and assignments that will challenge students, but also provide them with valuable experiences, especially real-life experiences [2]. According to [3], simulated real-world experience allows students to test their skills before leaving the educational environment.

Based on the current students' feedback on courses Internet Marketing (in the Business curriculum), and Web App Development (in the IT curriculum), students would like to have more real-world and real-time examples and as much as possible real-world experience [4]. The main purpose of the Internet marketing simulation is to provide both groups of students with course-specific knowledge and experience. In other words, the aim of Internet simulation is for students to develop a better understanding of theoretical fundamentals and course topics in a more interesting and dynamic way. Having in mind the collaborative nature of this project, the proposed simulation is enabling active learning in two courses within two different curriculums. The simulation supports the idea of topic delivery in a blended way. The lectures could be delivered face-to-face, while hands-on activities for both student sections could be delivered through the online Internet marketing simulation.

This learning by doing approach already exists in different Internet marketing simulations, such as Stukent Mimic Pro Simulation [5], Markstrat [6], or Anylogic [7], which we used as a benchmark to initially set parameters of our application. The main problem with other existing solutions is that they are not fully adaptable to all market dependencies, which our application is, and which is a prerequisite for fully online delivery of the courses

## II. Approach and Methods

The initial version of the Internet Marketing simulation comprised a simple business environment where students were supposed to manage digital marketing efforts for a hypothetical smartphone brand with a limited range of three devices with different technical specifications and targeted market. The simulation features an administration panel that provides the instructor with an option to modify various simulation parameters. Once the simulation is being set-up, the first round is a test round, which gives students an opportunity to get familiar with the interface. Both test and regular rounds consist of planning the budgets per digital platform and decisions about other parameters (e.g., keywords strategy). After each round, activity reports (i.e., digital platform insights/analytics) are generated providing students with valuable feedback on their actions. Additionally, students have an option to review market reports (containing competition's numbers and consumer preferences), but these reports would require spending of a certain amount of virtual money. Based on inputs from various reports, students have to manage digital marketing activities, such as content creation and content share/promotion via social media channels. The simulation gives students an option to decide how much of the virtual money should be spent per turn.

The simulation is designed as a three-step process:

*A. Planning stage*

In the planning stage, budgets, message strategy, keywords, and targeting options are set. Each student or team has to make a decision on budget allocation and spending per turn. The total simulation budget is predefined by the instructor. The weekly budget has to be split between products/brands and different communication channels/activities. The spending threshold per channel/activity is predefined (i.e., spending less than the threshold value will have no impact). Furthermore, more focused spending (e.g., limited to one product and/or fewer channels/activities) is more efficient, like in reality. This stage is completely developed and deployed in JavaScript technologies and is accessible through the front-end of our application.

*B. Execution of the plan*

This step is done on a daily or weekly turn, depending on the class delivery type. In the execution of the plan, students can make some low-scale changes to optimize the promotional effort. Each student or team has to select the keyword (or even multiple keywords, depending on the channel) per channel/activity. Keyword refers to the main topic communication is focused on (more details about the Keywords in section III B.). Keywords have a different impact on the efficiency of promotional activities depending on the product features, target audience preferences, and proposed budget. Synergy across channels can be achieved if communication is more focused (e.g., the same keyword is used across all communication channels). In other words, consistency is crucial. Students/teams are limited here with the drop-down menu.

In order to optimize the overall promotional effort, students can precisely define their target audience(s). Targeting primarily refers to the promotional activities and it is platform-specific. A proper combination of message strategy and targeting is essential for success. Appropriate targeting is also correlated with content creation and its quality score in the context of search marketing. In other words, a higher quality score, together with a consistent bidding strategy, will result in a more efficient search marketing campaign. This stage is completed on the basic level. At the moment, the basic principles of correlations have been implemented. However, as part of the future development, further and more detailed development of correlations between various planning variables is expected.

*C. Reporting*

Once the round/turn is finished, students receive distributed formal reports (digital platform insights/analytics) and some less formal (salesforce feedback). For the market report (containing competition's numbers and consumer preferences), students have to spend a certain amount of their overall marketing budget. This backend process is designed instead of one centralized database by using distributed Blockchain technologies, where we have developed reporting on the Blockchain technology. Initially, we encountered difficulties in payment options when tried to implement this in standard Ethereum Virtual Machine [8], which is part of the Ethereum network (this is to get the distribution of the reports, but also to prove the integrity of the data and to get an overall consensus of all of the parties in the reporting), so we plan to translate our solution to another Blockchain technology – HashNet network [9], where we can run our simulation reporting for free.

Our next step was adding the additional feature for the weekly budgeting that has to be split between products and different communication channels/activities to enable multi-user mode, so students (or teams of students) are able to compete with each other. Execution of the weekly plan based on weekly budgeting enables students to make some low-scale changes in order to optimize the promotional effort.

III. SYSTEM DESCRIPTION AND OUTCOMES

In this section, we will describe what are the goals of the simulation and how to interact with the simulation with three main tabs. The current simulation is providing the main features of Internet marketing principles, which could be easily extended upon further feedback.

*A. Goals of simulation*

From the perspectives of users, there are several goals in this simulation:

- The main goal is to increase sales for each of the three smartphone models;

- Additionally, students should increase brand loyalty/image because it will have a positive impact on sales in the next simulation turns (long-term perspectives).

Furthermore, Business students/teams are faced with the following set of goals of specific challenges:

- In order to increase sales, students have to appropriately allocate marketing budget and set-up all the correlated variables (e.g., keywords, post promotions).

- After each turn, students have to take into consideration outcomes from previous turns and adjust variables accordingly in order to optimize the performance of the digital marketing activities.

For IT students, goals are related to the technical aspects of the simulation. Therefore, IT students are dedicated to the collection of real-time inputs from the users for further application features development.

Students/teams have to manage digital marketing efforts for three product models (i.e., low-end, mid-end, and high-end products/brands). By default, the smartphone product category has been set-up. However, the instructor could make customization and manually change the product category. Since the marketing budget is limited, it is not possible to optimize promotional effort for all three products at the same time. Therefore, students have to make decisions about certain trade-offs. Ideally, only one model should be promoted within a turn (day or week) or over a period of few turns (if promoted for two or three weeks in a row, 20% better results should be expected). However, spending too much budget and focus on just one product can potentially have a negative impact on other models' sales (if promoted four weeks in a row, there is no additional impact; for the fifth week, there is a 25% penalty; for the sixth week 45% penalty and so on).

Additionally, the available budget would not support the full utilization of all channels within one turn. Therefore, students/teams needed to carefully develop their strategies and prioritize. Players should not try to utilize all channels at the same moment because the budget cannot support that.

*B. Simulation interface*

The first tab of the simulation is the planning stage. Each team will first have to decide how much money to spend in one turn. The total simulation budget is predefined. The weekly budget has to be split between products and different communication channels/activities, and an example is shown in Table I.

The spending threshold per channel/activity is also predefined, e.g., spending less than the threshold value will have no impact. Furthermore, more focused spending (e.g., limited to one product and/or fewer channels/activities) is more efficient. The numbers in the table are just for illustrational purposes. Students are responsible for all inputs.

TABLE I. BUDGET SPLIT PER CHANNEL/ACTIVITY IN COMMUNICATION PLATFORMS

| Communication platforms | Budget split per channel/activity | | | Budget per channel/activity |
|---|---|---|---|---|
| | *Phone 1* | *Phone 2* | *Phone 3* | *EUR* |
| Web site content | 50% | 40% | 10% | 500 |
| SEO | 60% | 30% | 10% | 200 |
| Facebook content production | 40% | 40% | 20% | 500 |
| Facebook page promotion | 60% | 30% | 10% | 100 |
| Facebook content promotion | 70% | 30% | 0% | 200 |
| Youtube content production | 100% | 0% | 0% | 500 |
| Youtube content promotion | 100% | 0% | 0% | 100 |
| Instagram content production | 50% | 40% | 10% | 300 |
| TOTAL | | | | 2400 |

The second tab of simulation is the message strategy described by one or more keywords. Here, students have to select a keyword (or even multiple keywords, depending on the channel) per channel/activity. This is the main topic of communication focus. Obviously, synergy across channels can be achieved if communication is more focused (e.g., one keyword is used in all channels). In other words, consistency is crucial. Students are limited here with the drop-down menu.

The current list of keywords for initial simulation consists of some simple points like Product features in general, Photography, Memory, Distinctive design, Practical design, Technical support reminder, Brand image related, Product differentiation, and Sales promotion support. Keywords have a different impact on the efficiency of promotional activities depending on the product features, target audience preferences, and proposed budget.

The last tab of simulation is focused on targeting. Targeting primarily refers to the promotional activities and it is platform-specific. A proper combination of message strategy and targeting is essential for success. For content creation, targeting will improve its quality score. The content quality score can improve promotional efficiency.

*C. Technologies*

Our system comprises several interconnected technologies that together create a useable, precise, and responsive system with agility for further implementations. One of the main goals was to focus on privacy guarantees for both students and professors such as decentralization and differential privacy. The system holds three main sectors that have explicit responsibilities and are interconnected by an intermediary level that manipulates and shares data.

Also, the key enforcer for using these technologies is the fact of the limited scope of the academic environment we were working in. Thus, we created a working prototype that uses the underlying technologies which allow free transactions per se, in comparison to EVM (Ethereum Virtual Machine) cost of transactions [8]. Nevertheless, the system can be migrated to any EVM-based blockchain network by just changing the targeted network and adjusting some communicational variables.

Blockchain supporting technologies:

- Truffle Suite – blockchain development environment
- Ganache – a local in-memory blockchain
- Web3.js – a web plug-in for Ethereum nodes
- MetaMask – Ethereum browser extension/crypto wallet
- Solidity – a programming language for smart contracts.

Truffle Suite [10] is a development environment that provides a set of tools that allow an easier development lifecycle for blockchain development. Also, it can be used as a testing framework and an asset pipeline for all blockchains targeting the Ethereum Virtual Machine (EVM). The environment offers built-in smart contract compilation, linking and deployment to multiple networks, interactive debugger, many libraries, and automated testing along with scriptable deployment and migrations frameworks, which allows accessible and straightforward implementation with different distributed ledger systems. From dependencies in our deployment, we were using Truffle v5.0.2 (core: v5.0.2) and truffle-contract v3.0.6.

Ganache is part of the Truffle Suite and presents a simulated personal blockchain environment replicating the behavior of the initial, real-world distributed ledgers. It can be used throughout the whole of the development cycle, allowing deployment and testing dApps in a safe but still analogous environment to the real concept. Ganache gives developers 10 mock accounts obtaining a certain amount of fake cryptocurrency, and with them, we were able to simulate a cohort of individual parties acting within the system. Each blockchain interaction on Ganache is saved with its TX hash, as well as the type of transaction, sender address, value (presented in given currency), gas used, gas price and limit with the block number in which the transaction was mined. For development we have used Ganache v2.4.0 with JSON Remote Procedure Call (RPC) server setup, the price of each unit of gas is set up at 20000000000 Wei, the gas limit is 6721975 units and the hard fork is set to Petersburg.

Web3.js is a set of libraries that allow communication with Ethereum nodes, both local and remotely placed ones via different network protocols. To connect Truffle Suite and Ganache to the browser, we have used web3.js v1.3.0 as a JavaScript web provider.

MetaMask allows websites to request Ethereum accounts, thus allowing them to operate Ethereum dApps. It does that by adding a provider object that indicates an Ethereum user, resulting in an Application Programming Interface (API) that can read data from the blockchains. In our system, we have used the MetaMask v4.0.2, which was a part of the user interface (UI) as well as a transaction communicator between the user and the simulation. Accordingly, when a distributed application wants to perform write a transaction on the blockchain, the user gets a secure interface to inspect the transaction before deciding to approve or decline it respectively. Consequently, our simulation asks the user on key system checkpoints to push through blockchain transactions. The setup of the system requires three central items which are detecting the Ethereum provider and Ethereum network on whom the user account is connected, and finally gets the user's Ethereum account.

Solidity is an object-oriented high-level language for the development of smart contracts. The language utilizes a simple single-slot database in which you can query and modify the code by calling the functions that manage the database. Our system uses solidity v0.5.0 with which we are deploying contracts to the Ganache in-memory representation of the Ethereum blockchain.

Lastly, from the technologies which are not directly connected to the blockchain reporting process, but are still a key part in the simulation, we have used HTML5, CSS, Bootstrap, JavaScript, and C# for the API.

The API business logic calculation subsystem uses C# with Microsoft (R) Build Engine version 16.2.32702+c4012a063 for the .NET Core version, and the project SDK Microsoft.NET.Sdk.Web with netcoreapp2.2 target framework. The API processes the largest amount of data and is the main sales force calculations mechanism. Through the simulation, JavaScript calls are made to the API via specialized controllers, which can handle user interaction with or without blockchain to make initial calculations, as well as finalize the weekly turn report.

*D. System Communication Flow*

At the start of the simulation, the admin enters pre-determined data, and this data is the is starting benchmark of globally defined values, which define the successfulness of each user's overall situation. These elements include the social media network's Number of likes, Post engagement, Page views, Average post reach, etc., and they are equally set for all users. This is the first interaction with the blockchain reporting mechanism, as transparency and immutability between all users of this data are imperative.

Secondly, the user starts defining Communication platform activities defined in Table I, and this data is gathered on the client side, and transformed into JavaScript Object Notation (JSON) data-interchange format. Then, the data is being pushed directly to the API at the beginning of a weekly turn, and cannot be changed anymore for that week, again on the next weekly turn, the student enters new choices. API immediately creates objects and calculates dependencies, which will be used to produce rudimentary figures which impact the core strategy results (platform insight analytics and salesforce feedback) and through the turn.

At the end of the weekly turn, the remaining user choice data regarding Brand content, Mobile content, Google ads, Facebook, Instagram, and YouTube ads keywords is being sent to API in order to complete analytics for the given turn. In parallel, the API takes global, blockchain-held information, that influences the current week's results. Afterward, the finalized weekly calculations from the API are combined with the blockchain saved global data in order to produce weekly progress reports, as well as present the total results at the end of the simulation.

*E. System Testing results*

Fig. 1 presents the total amount spend per user per simulation cycle. It is important to emphasize that Ethereum transaction fees variate constantly, and as Ethereum is the second-highest valued cryptocurrency, these fee costs are not insignificant for the user. The testing was conducted for the period from November 12, 2020, to November 17, 2020.

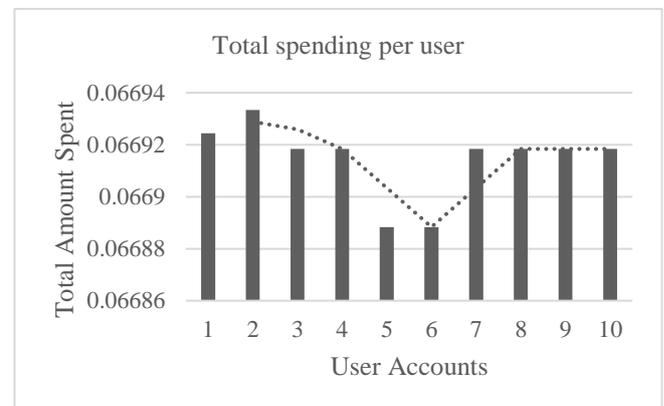

Fig. 1. Preliminary user spending per simulation cycle results

The first four user accounts were tested within a space of a couple of hours, so the result prices fluctuate slightly. Fig. 1 presents data from Blockchair, a tool that provides different data regarding Ethereum variables [8]. The slight fall in the transaction costs of the user accounts 5 and 6 can be seen in Fig. 2. As of 14th, the average daily price was 0.00259 ETH and on the 15th 0.00241 ETH, as compared to previous and after days when it circulated around 0.0032 ETH. Finally, the last four user accounts were tested successively, with the time between transactions being around 30 seconds.

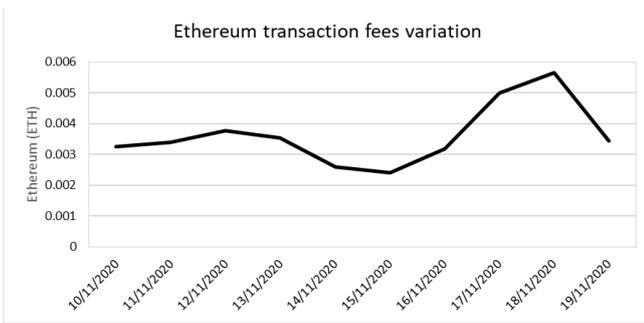

Fig. 2. Ethereum Average transaction fee variation

Fig. 3 presents the average time per user's transaction. As it can be seen, time varies between 3 and 6 seconds per transaction, which is within some average time needed for transactions in Ethereum [8]. From the perspective of simulation reporting, it shows that reports are quickly distributed among the students/teams, and in this way, they have a distributed reporting system that can be proved by each team (which is important from the integrity perspective). Time to the finality of a transaction by using the Ethereum network could be the problem for cryptocurrency application but for proving the integrity of report transactions in our simulation is not imperative. In this way, students can run the simulation in multi-user mode to compete with each other and yet to get consistent reporting, which integrity is proved and shared by the complete network.

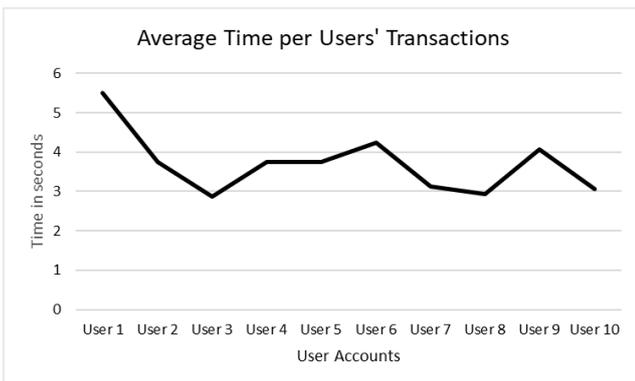

Fig. 3. Average time per all user's transactions

## IV. CONCLUSIONS AND SUMMARY

Our system comprises several interconnected technologies that together create a useable, precise, and responsive system with agility for further operations. We covered Internet marketing principles, distributed Web application development based on Blockchain reporting. One of the main goals was to focus on privacy guarantees for both students and professors, such as decentralization and differential privacy. The system holds three main sectors which have explicit responsibilities and are interconnected by an intermediary level that manipulates and shares data.

Our application provides a high level of customization through the admin tools accessible by the instructors, but also through the user panel and different options based on the distributed Blockchain reports. Also, our application can run in multi-user mode, so students (or teams of students) are able to compete with each other. This enables the gamification of the learning process, which is one of the basics for active learning, one of the pillars of our transition to digital delivery of teaching and learning activities. International Business students get a better understanding of the Internet marketing principles not only by using this application, but also by partly designing the business logic (other parts are on the instructor on this course), so they will construct the knowledge and understanding. IT students build the real-world application (and not usually some in-class application no one uses after they complete the course), will be able to interact with the real users (in this case online with their counterpart International Business students) about the user experience of their application (usually they are able just to get instructor's feedback, which is more theoretical). Both boost the way how they construct their knowledge and understanding. With this overall simulation approach their later probability of failure in their businesses will be lower. This option is also tightly related to changes in IT technologies for real-world applications and online course deliveries and we used this COVID-19 outbreak to boost the capabilities and competencies of our international students that are now online.

ACKNOWLEDGMENT

This work is supported by the RIT PLIG grant.